\documentclass[twocolumn]{aastex6}

\fullcollaborationName{}

\shorttitle{Multiple rings in the disk of GM Aur}
\shortauthors{Mac\'{\i}as et al.}

\begin{document}


\title{Multiple rings in the transitional disk of GM Aurigae revealed by VLA and ALMA}


\author{Enrique Mac\'{\i}as, Catherine C. Espaillat, \'Alvaro Ribas}
\affil{Department of Astronomy, Boston University, 725 Commonwealth Avenue, Boston, MA 02215, USA
{\tt emacias@bu.edu}}

\author{Kamber R. Schwarz}
\affil{Department of Astronomy, University of Michigan, 1085 South University Avenue, Ann Arbor, MI4810, USA}

\author{Guillem Anglada, Mayra Osorio}
\affil{Instituto de Astrof\'\i sica de Andaluc\'\i a (CSIC) Glorieta de la Astronom\'\i a s/n E-18008 Granada, Spain}

\author{Carlos Carrasco-Gonz\'alez}
\affil{Instituto de Radioastronom\'{\i}a y Astrof\'{\i}sica 
UNAM, Apartado Postal 3-72 (Xangari), 58089 Morelia, Michoac\'an, Mexico}

\author{Jos\'e  F. G\'omez}
\affil{Instituto de Astrof\'\i sica de Andaluc\'\i a (CSIC) Glorieta de la Astronom\'\i a s/n E-18008 Granada, Spain}

\and

\author{Connor Robinson}
\affil{Department of Astronomy, Boston University, 725 Commonwealth Avenue, Boston, MA 02215, USA}

\begin{abstract}
Our understanding of protoplanetary disks is rapidly departing from the classical view of a smooth, axisymmetric disk. This is in part thanks to the high angular resolution that (sub)mm observations can provide. Here we present the combined results of ALMA (0.9 mm) and VLA (7 mm) dust continuum observations toward the protoplanetary disk around the solar analogue GM Aur. Both images clearly resolve the $\sim$35 au inner cavity. The ALMA observations also reveal a fainter disk that extends up to $\sim250$ au. We model our observations using two approaches: an analytical fit to the observed deprojected visibilities, and a physical disk model that fits the SED as well as the VLA and ALMA observations. Despite not being evident in the deconvolved images, the VLA and ALMA visibilities can only be fitted with two bright rings of radii $\sim$40 and $\sim$80 au. Our physical model indicates that this morphology is the result of an accumulation or trapping of large dust grains, probably due to the presence of two pressure bumps in the disk. Even though alternative mechanisms cannot be discarded, the multiple rings suggest that forming planets may have cleared at least two gaps in the disk. Finally, our analysis suggests that the inner cavity might display different sizes at 0.9 mm and 7 mm. This discrepancy could be caused by the presence of free-free emission close to the star at 7 mm, or by a more compact accumulation of the large dust grains at the edge of the cavity. 
\end{abstract}

\keywords{protoplanetary disks --- planet-disk interactions --- stars: individual (GM Aurigae) --- stars: pre-main sequence --- techniques: interferometric}



\section{Introduction} \label{sec:intro}

Planetary systems have long been known to form in protoplanetary disks. These objects were usually assumed to be smooth, axisymmetric structures. However, this paradigm has recently started to change; complex structures such as gaps, rings, vortices, or spiral arms are being imaged and resolved using the high angular resolution provided by current radiointerferometers, such as the Atacama Large Millimeter/submillimeter Array (ALMA; \citealp{alm15,and16,per16,ise16}) and the NSF's Karl G. Jansky Very Large Array (VLA; \citealp{oso14,car16,mac16}).

The discovery of these disk substructures has sparked a lot of interest, partly because they could solve one of the fundamental problems of the core-accretion model, our current paradigm of planetary formation. According to this scenario, planets form through the sequential growth and aggregation of dust particles (\citealp{tes14}, and references therein). However, large grains in the disk suffer a gas drag force that tends to drift the particles toward the higher pressures found in the inner regions of the disk \citep{whi72}. Dust evolution models predict that the mm/cm-sized particles in the disk should drift very rapidly, hence falling onto the star before they can grow up to planetesimal sizes \citep{bra07}. Local maxima in the gas pressure of the disk can provide a solution for this problem since they can stop the migration of large particles, accumulating them and allowing them to grow even more efficiently \citep{chi10,pin12}. The disk substructures revealed by VLA and ALMA could be the result of these dust accumulations.

Among the several physical processes that have been proposed to explain the observed disk substructures, non-smooth gas distributions, resulting in the onset of dust traps in the disk, are often involved. Some examples of these scenarios are the dynamic interactions with planets \citep{zhu14,bae17}, the magneto-rotational instabilities (MRI) in magnetized disks \citep{flo15}, or the back-reaction of the dust onto the gas \citep{gon15,gon17}. However, other physical processes, such as changes in dust properties related to condensation fronts or snowlines of volatiles \citep{oku16,pin17b}, can produce similar features without the need of pressure bumps. More observations are needed to better understand the commonality of disk substructures and their role in the dust evolution and planetary formation process.

Here we focus on the transitional disk around the T Tauri star GM Aur, a young solar analogue (K5.5 spectral type, $L_{\star}\simeq0.9$ $L_{\odot}$, $M_{\star}\simeq1.1~M_{\odot}$; \citealp{ken95,esp11}) located at $160\pm2$ pc \citep{gai16,gai18} in the Taurus-Auriga cloud.
The disk of GM Aur has been very well studied from UV to cm wavelengths.
From SED modeling, it was shown to present an inner cavity in the optically thick dust \citep{cal05}, although filled with residual optically thin small particles \citep{esp10,esp11}. Furthermore, the star of GM Aur presents significant accretion of material (0.39--1.1$\times10^{-8}~M_{\odot}$ yr$^{-1}$; \citealp{ing13,ing15}), which suggests that the inner cavity is also partially filled with gas. Sub-mm observations later confirmed the presence of a disk cavity of $20-30$ au in radius, but could not unveil additional details of the disk due to the limited sensitivity and angular resolution \citep{hug09,and11}. Recently, VLA observations at cm wavelengths revealed that GM Aur contains a substantial amount of ionized gas, suggesting that the disk could be undergoing some degree of photoevaporation \citep{mac16}. Some studies have shown that photoevaporation can explain the presence of small, significantly depleted cavities \citep{ale14,owe16}. However, 
the size of GM Aur's cavity, combined with the presence of gas and small dust grains inside the cavity, points toward a dynamical clearing origin \citep{esp14}, indicating that planetary formation is probably taking place in GM Aur.

In this paper we present new mm observations of the protoplanetary disk of GM Aur obtained with ALMA at 0.9 mm and with the VLA at 7 mm. We model the observations and find that the large dust grains distribution of the disk is mainly composed of two rings, where large dust grains are being accumulated, surrounded by a fainter and more extended outer disk that has been significantly depleted from large dust grains due to radial drift.

\section{Observations} \label{sec:observations}

\subsection{ALMA Observations} \label{observations:ALMA}

The ALMA observations were performed at Band 7 ($\sim0.9$ mm) during Cycle 4 (PI: Kamber Schwarz). The data were obtained on 2016 November 11 using 39 antennas of the 12 m array, with baselines ranging from $\sim15$ m to $\sim850$ m. 

The correlator was set to use three different spectral windows. One was centered at 317.989 GHz and was dedicated exclusively to detect the continuum dust emission, using the highest bandwidth possible (2 GHz). The other two spectral windows were tuned to observe the $^{13}$CO ($3-2$), and C$^{18}$O ($3-2$) molecular transitions with rest frequencies 330.588 GHz and 329.331 GHz, respectively. Both lines were observed with a spectral resolution of $\sim122$ kHz, covering a total bandwidth of $\sim234.4$ MHz in each spectral window. In this paper we present the results of the dust continuum emission, while the molecular emission will be analyzed in Schwarz et al. (in prep.).

Data calibration and editing were performed using the data reduction package CASA (Common Astronomy Software Applications; version 4.7.0)\footnote{https://science.nrao.edu/facilities/vla/data-processing}. The quasar J0522-3627 was observed as the absolute flux calibrator, with an expected uncertainty in the flux scale of $\sim10\%$. Bandpass and complex gain calibration were performed using J0510+1800. The pipeline was used to calibrate the data, with no additional flagging required. Self-calibration was applied on GM Aur, which successfully improved the quality and signal-to-noise ratio of the data. 

We used the task CLEAN of CASA (version 4.5.3) to produce cleaned images from the calibrated visibilities. The 2 GHz spectral window was combined with the line-free channels of the two 234 MHz spectral windows to maximize the continuum sensitivity. In order to take into account the frequency dependence of the emission, the multi-frequency deconvolution algorithm built in CLEAN was used with nterms=2 \citep{con90}. The resulting central frequency of the images was 323.84 GHz ($\sim0.93$ mm). 

In addition, different visibility weightings were applied to obtain images with different angular resolutions. The top-left panel of Fig. \ref{fig:maps} shows the image obtained with a robust parameter of 0.5, which gives a good compromise between sensitivity and angular resolution. This resulted in a synthesized beam size of $0\rlap.''39\times0\rlap.''28$ (PA$=0^{\circ}$) and an rms of $0.10$mJy beam$^{-1}$, producing a peak SNR of 500. An image obtained with uniform weighting is shown in the top-right panel of Fig. \ref{fig:maps}. The resulting syntesized beam size is $0\rlap.''29\times0\rlap.''20$ (PA$=14^{\circ}$), with an rms of $0.35$ mJy beam$^{-1}$ and a peak SNR of 92. This image has a significantly higher angular resolution but it is less sensitive to the extended emission.

\subsection{VLA Observations} \label{observations:VLA}

We obtained Q band ($\sim7$ mm) data of GM Aur using the VLA of the National Radio Astronomy Observatory (NRAO)\footnote{The NRAO is a facility of the National Science Foundation operated under cooperative agreement by Associated Universities, Inc.}. The observations were performed on 2016 June 17 and July 20 using the B configuration, with antenna separations that ranged from $\sim150$ m to $\sim10000$ m. In order to maximize the continuum sensitivity, the WIDAR correlator was set to use the 3-bit sampler together with a total bandwidth of 8 GHz. The quasar 3C147 was used as the flux and bandpass calibrator, whereas J0438+3004 was observed to perform the complex gain calibration. The expected uncertainty in the absolute flux calibration is $\sim10~\%$.

The observations were calibrated using the VLA calibration pipeline within CASA (version 4.7.2). After processing the data through the pipeline, calibrated data were inspected, flagged, and then rerun through the pipeline again. Overall, a small amount of flagging was needed for a few baselines or antennas at particular spectral windows and channels. For the observations on July 20, the antenna \textit{ea25} was problematic and had to be flagged.

\begin{figure*}
\figurenum{1}
\plotone{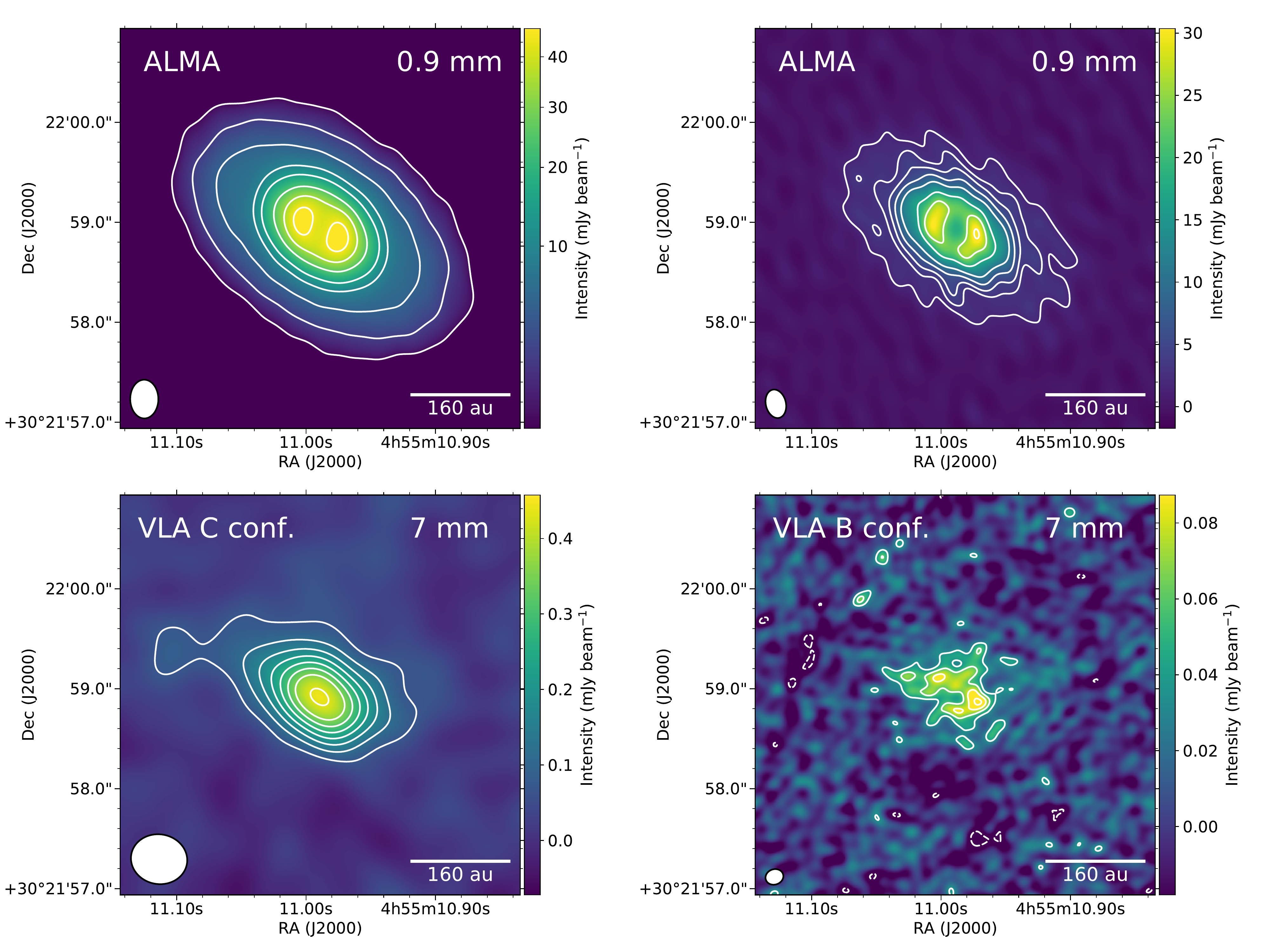}
\label{fig:maps}
\caption{Observed images of the mm emission of GM Aur. \textit{Top-left}: ALMA image at 0.9 mm, obtained using Briggs weighting with robust=0.5 (synthesized beam=$0\rlap.''39\times0\rlap.''28$, PA=$0^{\circ}$). Contour levels are 5, 20, 50, 100, 150, 250, 350, and 450 times the rms of the map, $0.10$ mJy beam$^{-1}$. \textit{Top-right}: ALMA image at 0.9 mm, obtained using uniform weighting (synthesized beam=$0\rlap.''29\times0\rlap.''20$, PA=$14^{\circ}$). Contour levels are 5, 9, 13, 20, 30, 50, 70, and 90 times the rms of the map, $0.35$ mJy beam$^{-1}$. \textit{Bottom-left}: Image at 7 mm, obtained using the C configuration of the VLA and natural weighting (synthesized beam=$0\rlap.''58\times0\rlap.''50$, PA=$81^{\circ}$), adapted from \citet{mac16}. Contour levels are 3, 5, 7, 9, 11, 13, 15, and 18 times the rms of the map, $24~\mu$Jy beam$^{-1}$. \textit{Bottom-right}: Image at 7 mm, obtained using the B configuration of the VLA and natural weighting (synthesized beam=$0\rlap.''18\times0\rlap.''15$, PA=$-72^{\circ}$). Contour levels are $-$3, 3, 5, 7, 9, and 11 times the rms of the map, $12~\mu$Jy beam$^{-1}$.}
\end{figure*}

Cleaned images were obtained with the CLEAN task of CASA (version 4.5.3), using the multi-frequency deconvolution algorithm as in \S\ref{observations:ALMA}. The frequency center of the images was 44.0 GHz ($\sim6.8$ mm). We applied a natural weighting to maximize the signal-to-noise ratio of the image, which provided us with an angular resolution of $0\rlap.''18\times0\rlap.''15$ (PA$=-72^{\circ}$), higher than that of the ALMA observations, and an rms of $12~\mu$Jy beam$^{-1}$, resulting in a peak SNR of 9.75. The resulting image is plotted at the bottom-right panel of Fig. \ref{fig:maps}. As a comparison, at the bottom-left panel we show a lower angular resolution image at 7 mm obtained with C configuration, previously reported in \citet{mac16}.

\section{Results} \label{sec:results}

Our ALMA (0.9 mm) and VLA (7 mm) images reveal, with a high degree of detail, the dust thermal emission of the protoplanetary disk of GM Aur. 
In order to analyze the dust emission and the possible differences that arise at these two wavelengths, we compare the deconvolved images as well as the averaged radial intensity profiles.

\begin{figure*}
\figurenum{2}
\plotone{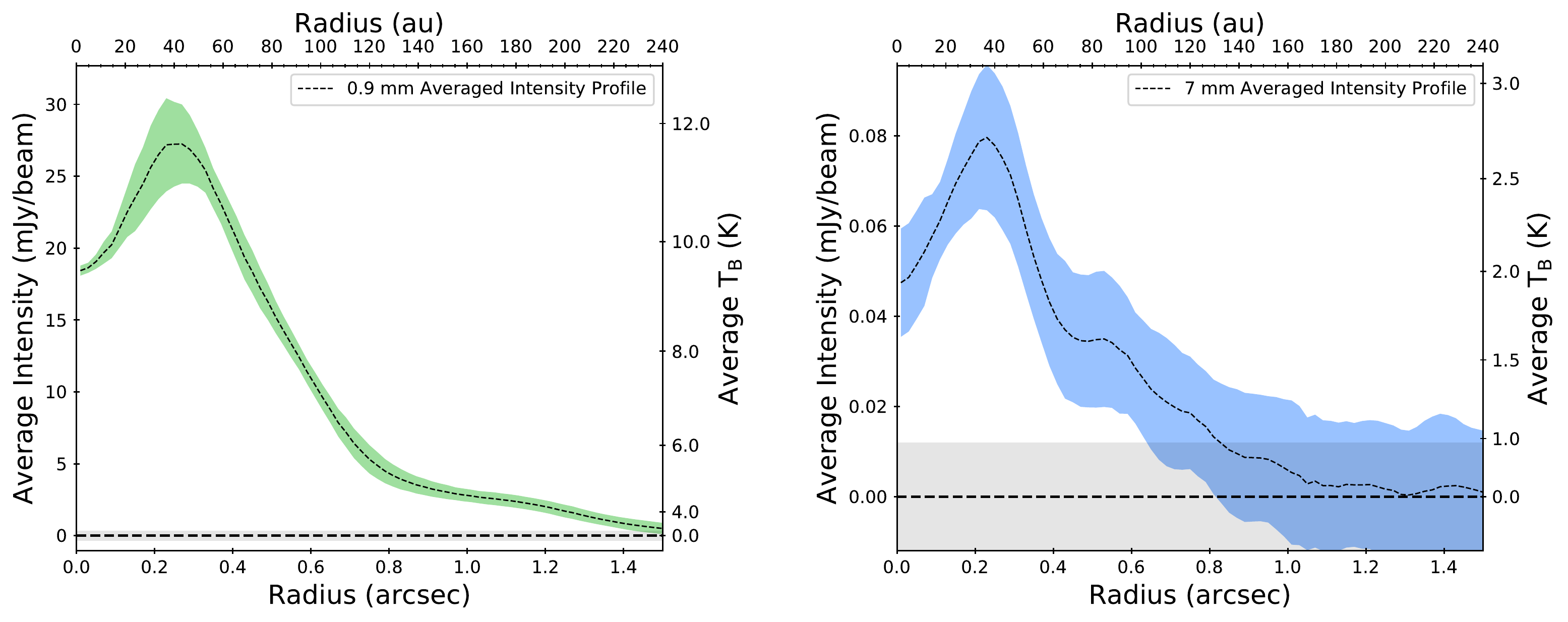}
\label{fig:radprofile}
\caption{Averaged radial intensity profiles of the 0.9 mm (left) and 7 mm (right) emission of GM Aur. These profiles have been obtained from the uniform-weighted image at 0.9 mm (top-right panel in Fig. \ref{fig:maps}), and from the B configuration image at 7 mm (bottom-right panel in Fig. \ref{fig:maps}), respectively. The colored region indicates the standard deviation at each radius, which can increase because of the noise in the image, the effects of the elongated beam, or genuine departures from axi-symmetry. The horizontal dashed lines mark the y=0 level. The grey areas surrounding these lines show the 1 $\sigma$ (rms) level.}
\end{figure*}

We compute the radial profiles by averaging the emission in concentric ellipses, taking into account the inclination and position angle of the disk ($i=52.8^{\circ}$, PA$=56.5^{\circ}$; see \S\ref{subsec:SBPmodel}). The width of the ellipses was set to $0\rlap.''02$, a relatively small fraction of the beam size. The results are shown in Fig.\ref{fig:radprofile}.

Both ALMA images (top panels of Fig. \ref{fig:maps}) show a similar morphology, with a bright compact ring surrounding a cavity of $\sim35$ au, and a fainter disk  au. The robust-weighted image better recovers the extended emission and shows the high contrast in surface brightness between the compact ring and the extended disk. These two components can also be clearly seen in the radial intensity profile (left panel of Fig. \ref{fig:radprofile}). By enclosing in a box the emission above $3 \sigma$ in the robust-weighted image, we measure a flux density of $380\pm40$ mJy\footnote{The uncertainty in the measured flux density combines the absolute flux calibration uncertainty and the uncertainty introduced by the rms noise. The latter is computed as the rms noise times the square root of the area of the box in number of beams.}. A similar box enclosing the uniform-weighted image yields a flux density of $350\pm40$ mJy. We note that the uncertainties in the flux densities of both images are completely dominated by the systematic uncertainty introduced by the absolute flux calibration. The difference in 30 mJy is hence significant, indicating that the uniform-weighted image is missing $\sim8\%$ of the emission of the disk. In comparison, previous single dish submm observations measured a flux density of $730\pm70$ mJy at 0.80 mm \citep{wei89}. Assuming a spectral index of 3.05, as measured by \citet{mac16}, the single dish flux density at $\sim0.93$ mm would be $470\pm50$. Although being almost within error bars, this value is slightly higher than the flux in the robust-weighted image, suggesting that some extended emission might not be recovered by the ALMA observations.

The size of the cavity is consistent with previous SED modeling \citep{esp11} and submm observations of GM Aur \citep{and11}. However, these previous observations lacked the sensitivity to detect the extended component. While the extended disk emission in the ALMA observations appears very axisymmetric, our uniform-weighted image suggests that the compact ring of emission could be slightly azimuthally asymmetric. The peak intensity of the western side is $\sim5\sigma$ brighter than the eastern side, which represents a $\sim5.6\%$ difference. Furthermore, the intensity peak is located at a position angle of $\sim270^{\circ}$, displaced $\sim20^{\circ}$ north from the major axis of the extended disk.

On the other hand, the VLA observations (bottom panels of Fig. \ref{fig:maps}) appear to show a more compact morphology. Our new B configuration observations significantly improve the angular resolution of the previous C configuration data \citep{mac16}. The 7 mm emission shows a ring-like morphology that surrounds a cavity of $\sim30$ au and extends up to $\sim120$ au. We measure a flux density of $1.36\pm0.15$ mJy, which is consistent with the previous C configuration observations \citep{mac16}. 

Interestingly, a small bump is seen in the 7 mm radial intensity profile at $\sim80$ au (right panel of Fig.\ref{fig:radprofile}). This feature resembles the shape produced by a ring in the disk, but the low signal-to-noise ratio of the image prevents us from confirming its presence. We note that the angular resolution of the ALMA observations is too low to discern if this feature is also present in the 0.9 mm radial profile. 
A detailed modeling of the interferometric visibilites at 0.9 mm and 7 mm confirms the presence of at least two rings in the disk (see \S\ref{sec:modeling}).

The disk cavity radius appears to be smaller at 7 mm than at 0.9 mm, but a more detailed analysis of the observations is needed to discern whether this discrepancy is significant, given the difference in angular resolution between the VLA and the ALMA observations (see \S\ref{subsec:cavity}). Furthermore, the ring seems to be partially lopsided in our B configuration image, with the western side of the disk being $\sim26\%$ ($\sim2\sigma$) brighter than the eastern side. An asymmetry with such a low signal-to-noise ratio could be created by a noise peak or a cleaning artifact. However, its position is coincident with the asymmetry detected at 0.9 mm, suggesting that the emission of the ring could be indeed slightly asymmetric in azimuth. 
More sensitive observations are needed to confirm its presence.

Another difference between the ALMA and the VLA observations is the presence of the extended disk emission at 0.9 mm. The fact that this extended component is not evident in the VLA images could be the result of differences in the radial distribution of the dust particles traced at the two different wavelengths.
However, both the C configuration and B configuration images of the 7 mm emission are probably sensitivity limited, so a more cautious analysis is needed to draw any robust conclusions. In any case, the high contrast in surface brightness between the extended disk and the compact ring shown at 0.9 mm suggests that the outer regions of the disk are indeed significantly depleted of large dust grains (see \S\ref{subsec:DIADmodel}). 

It is important to note that the CLEAN algorithm used to obtain the images shown in Fig. \ref{fig:maps} is just one of the different deconvolution methods (albeit the most commonly used) that can be applied to interferometric visibilities in order to obtain the intensity distribution of the source. Despite the flexibility and effectiveness of this method, it usually results in images that lack information on the highest spatial frequencies sampled by the longer baselines of the interferometer \citep{zha16}. At the same time, a coarse sampling of the uv plane can result in artifacts being introduced during the cleaning process. Directly analyzing the observed visibilities can provide us with additional information while also avoiding these possible artifacts. In particular, if we assume that the disk emission is intrinsically axisymmetric, we can use the deprojected visibility profile to better analyze the radial structure of the emission (e.g., see \citealp{hug07}). We follow this approach and model the observed visibilities in \S\ref{sec:modeling}.

\begin{figure*}
\figurenum{3}
\plotone{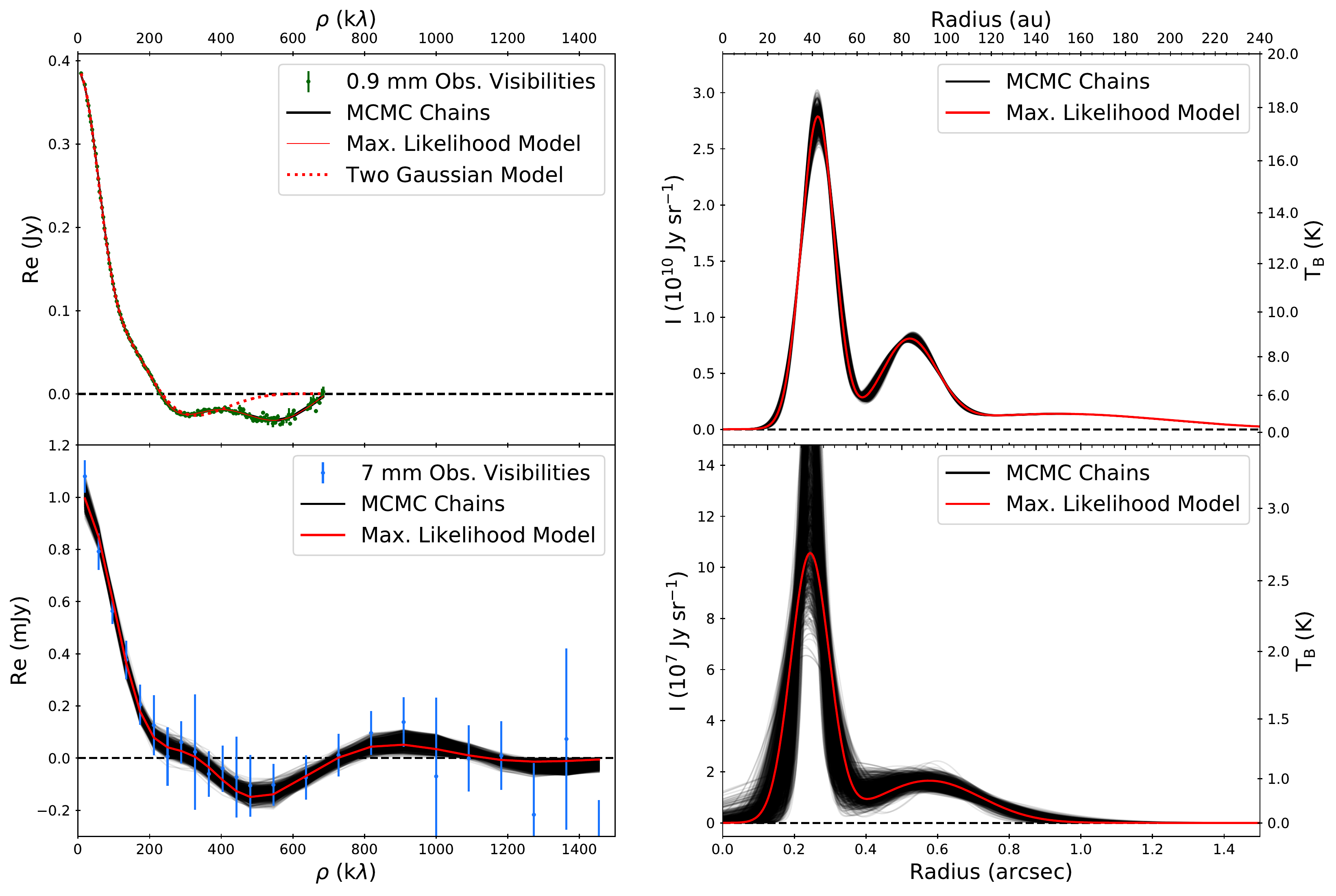}
\label{fig:vismodel}
\caption{Results of our analytical surface brightness profile model of GM Aur at 0.9 mm (top panels) and 7 mm (bottom panels). The left panels show the real part of the deprojected interferometric visibilities, while the right panels show the corresponding surface brigthness profiles. The observed visibilities are plotted in the left panels as dots with error bars. Even though we fit each observed point in the uv plane, we plot the binned visibilities in ranges of deprojected uv-distances to facilitate the readability of the figure. Black lines show 1000 random models selected from the last 5000 chains in our MCMC runs. The MCMC chain with the maximum likelihood is plotted with a red line. We are able to reproduce the observed visibilities using three Gaussian rings at 0.9 mm, and two Gaussian rings at 7 mm. The best fit model to the 0.9 mm visibilities using only two Gaussian rings is shown as a red dotted line. The horizontal dashed lines indicate the y=0 level.}
\end{figure*}

\section{Modeling}\label{sec:modeling}

In the following section we further analyze our VLA and ALMA data by modeling the interferometric visibilities and surface brightness profile of GM Aur. We follow two different approaches. First, we use an analytical model of the surface brightness profile of the disk to directly fit the deprojected visibilities. We then use the information obtained from this simple model to inform a more complete physical model that can reproduce both the Spectral Energy Distribution (SED) of GM Aur as well as the VLA and ALMA observations.

\subsection{Surface Brightness Profile Model}\label{subsec:SBPmodel}

We aim at analyzing the structure of the mm emission of GM Aur in more detail by directly fitting an analytical model of the surface brightness profile of the disk to the real part of deprojected visibilities. This \textit{semi-empirical} approach has the advantage that it will use the whole extent of the uv coverage, giving us more information on the highest spatial frequencies than a deconvolved image. At the same time, this modeling does not require any physical assumptions for the disk.
We note, however, that this approach is still not model-independent, since we need to assume a particular functionality for the surface brightness profile.  

Based on the multi-ringed structure observed in several protoplanetary disks (e.g., \citealp{zha16,and16,mac17}), we assume a relatively simple intensity profile with a set of Gaussian rings centered at different positions:
\begin{equation}
    I(\theta)=\sum_{i}^{N} a_i ~\textrm{exp} \Big( -\frac{(\theta-x_i^2)}{2\sigma_i^2} \Big),
\end{equation}
where $N$ is the number of Gaussian rings included in the model, $\theta$ is the radial angular scale of the disk, and $a_i$, $\sigma_i$, and $x_i$ are the three free parameters (scaling factor, standard deviation, and center position) for each Gaussian ring. If we assume that the emission is axisymmetric, the visibility profile as a function of deprojected uv-distance can be calculated using a Hankel transform \citep{pea99}:
\begin{equation}
    V(\rho)=2\pi\int_{0}^{\infty} I(\theta) \theta J_0(2\pi\rho\theta) d\theta, \\
\end{equation}
\begin{equation}
    \rho=\sqrt{u'^2+v'^2},
\end{equation}
\begin{equation}
u'=\cos{i}\times(u\cos{\textrm{PA}}-v\sin{\textrm{PA}}),
\end{equation}
\begin{equation}
 v'=u\sin{\textrm{PA}} + v\cos{\textrm{PA}}
\end{equation}

where $\rho$ is the deprojected uv-distance (in units of $\lambda$), $J_0$ is the zeroth-order Bessel function of the first kind, $i$ is the inclination, and PA is the position angle of the disk. The latter two are introduced as free parameters in our model, together with a potential offset in RA ($\Delta$RA) and DEC ($\Delta$DEC) from the position of the central star, where the phase center of the observed visibilities was initially set. The stellar position was obtained from the position and proper motions measured by Gaia (RA = 04h55m10.9885s DEC = 30d21m58.947s for the ALMA observations, RA = 04h55m10.9884s DEC = 30d21m58.957s for the VLA observations; \citealp{gai18}). A similar method has been successfully used to analyze ALMA observations of other protoplanetary disks \citep{zha16,pin17a}. 

In summary, the free parameters of our model are the inclination, PA, offsets in RA and DEC, and $a_i$, $\sigma_i$, and $x_i$ for each Gaussian ring.  
We use each point in the uv plane to fit our model. We follow a Bayesian approach and use the Markov Chain Monte Carlo (MCMC) ensemble samplers with affine invariance \citep{goo10} provided with the \texttt{EMCEE} package \citep{for13}. Details about the initial positions of the chains and the priors can be found in Appendix \ref{appendix:MCMC}.

Based on the shape of our deconvolved 0.9 mm image, we first used two Gaussians to fit the ALMA observations. Our model managed to fit the shorter visibilities, but it failed to reproduce the second minimum in the negative visibilities at $\sim600$ k$\lambda$ (see left panels in Fig. \ref{fig:vismodel}). Therefore, we added a third Gaussian to the model. In this way, we managed to obtain a good fit to the ALMA visibilities, resulting in a double-ring morphology surrounded by a flatter third ring. Nevertheless, the third Gaussian proved to be unnecessary for the VLA data, as a good fit was achieved with only two Gaussians for the compact double ring. The analytical model was not able to find a third component similar to the extended disk found in the ALMA data, probably due to the lower signal-to-noise ratio of our VLA data. Thus, we chose to use only two Gaussians to fit the VLA visibilities.

The results of our surface brightness profile fit can be seen in Fig. \ref{fig:vismodel}. The corner plots are shown in Appendix \ref{appendix:MCMC}. The medians, as well as the 16th and 84th percentiles, of the posterior distributions of the model parameters are shown in Table \ref{Tab:SBPmodel}. 

The visibility profiles of both the VLA and ALMA observations can only be reproduced with a disk composed of two rings of emission at $\sim40$ and $\sim80$ au, as well as a third extended component for the ALMA data. The double ring morphology was in fact suggested by the averaged radial profile of the 7 mm cleaned image (see Fig. \ref{fig:radprofile}). The fit to the ALMA observations also finds a more extended disk component that reaches $\sim250$ au, as was indicated by the cleaned images (see Fig.\ref{fig:maps}). Furthermore, the size of the disk cavity appears to be smaller at 7 mm ($x_1=39\pm1$ au) than at 0.9 mm ($x_1=42.5\pm0.2$ au), as initially suggested by the averaged radial profiles (Fig.\ref{fig:radprofile}). We discuss this further in \S\ref{subsec:cavity}.

Additionally, we find an inclination of $52.77^{\circ} ~^{+0.05}_{-0.04}$ (face-on: $i=0^{\circ}$, edge-on: $i=90^{\circ}$) and a PA of $56.45^{\circ}~^{+0.06}_{-0.05} $. Interestingly, previous studies found different position angles of the disk for the dust emission ($64^{\circ}$; \citealp{hug09,and11}) and the CO emission ($51^{\circ}$; \citealp{dut98}). Here we find a value that is in between both estimates, although slightly closer to the estimate from the CO emission. On the other hand, our estimated inclination is very close to the previously measured inclination of $55^{\circ}$ \citep{hug09,and11}.


\floattable
\begin{deluxetable}{cccc}[ht!]
\tablecaption{Results of Surface Brightness Profile Model (medians, $16\%$, and $84\%$ percentiles). The first four parameters are the inclination ($i$), position angle (PA), offset in RA ($\Delta$RA), and offset in DEC ( $\Delta$DEC). The rest of the parameters represent the scale factor ($a_i$), standard deviation ($\sigma_i$), and position ($x_i$) of each Gaussian ring. \label{Tab:SBPmodel}}
\tablehead{
\colhead{ }  & \colhead{0.9 mm} & & \colhead{7 mm}   }
\startdata
$i (^{\circ})$ & 52.77$^{+0.05}_{-0.04}$ & & 52.77$^{+0.05}_{-0.04}$~\tablenotemark{a} \\
PA $(^{\circ})$ & 56.45$^{+0.06}_{-0.05}$ & & 56.45$^{+0.05}_{-0.05}$~\tablenotemark{a} \\
$\Delta$RA  (miliasec) & -31.1$^{+1.9}_{-1.8}$ & & 10$^{+3}_{-3}$ \\
$\Delta$DEC  (miliasec) & 43.5$^{+1.7}_{-1.8}$ & & -6$^{+3}_{-4}$ \\  
\hline
$a_{1}$ (Jy sr$^{-1}$) & $2.74^{+0.09}_{-0.08}\times10^{10}$ & & $1.1^{+0.4}_{-0.2}\times10^{8}$ \\
$\sigma_{1}$ & $0\rlap.''047^{+0\rlap.''002}_{-0\rlap.''002}$ ($7.5^{+0.3}_{-0.3}$ au) & & $0\rlap.''044^{+0\rlap.''014}_{-0\rlap.''014}$ ($7^{+2}_{-2}$ au)  \\
$x_{1}$      & $0\rlap.''2657^{+0\rlap.''0014}_{-0\rlap.''0014}$ ($42.5^{+0.2}_{-0.2}$ au)  & & $0\rlap.''243^{+0\rlap.''006}_{-0\rlap.''007}$ ($39^{+1}_{-1}$ au)  \\
\hline
$a_{2}$ (Jy sr$^{-1}$)     & $7.54^{+0.16}_{-0.13}\times10^{9}$ & & $1.7^{+0.2}_{-0.2}\times10^{7}$ \\
$\sigma_{2}$ & $0\rlap.''077^{+0\rlap.''004}_{-0\rlap.''004}$ ($12.3^{+0.6}_{-0.6}$ au) & & $0\rlap.''17^{+0\rlap.''05}_{-0\rlap.''04}$ ($27^{+8}_{-6}$ au) \\
$x_{2}$      & $0\rlap.''521^{+0\rlap.''004}_{-0\rlap.''004}$ ($83.4^{+0.6}_{-0.6}$ au) & & $0\rlap.''55^{+0\rlap.''04}_{-0\rlap.''07}$ ($88^{+6}_{-11}$ au) \\
\hline
$a_{3}$ (Jy sr$^{-1}$)     & $1.396^{+0.008}_{-0.008}\times10^{9}$ & & -- \\
$\sigma_{3}$ & $0\rlap.''308^{+0\rlap.''003}_{-0\rlap.''003}$ ($49.3^{+0.5}_{-0.5}$ au) & & --  \\
$x_{3}$      & $0\rlap.''934^{+0\rlap.''006}_{-0\rlap.''006}$ ($149^{+1}_{-1}$ au) & & --  \\
\enddata
\tablenotetext{a}{The ALMA posteriors of the inclination and PA were used as priors for the VLA observations. }
\end{deluxetable}

\subsection{Physical Disk Model}\label{subsec:DIADmodel}

In order to obtain more information about the physical properties of the disk of GM Aur, we aim at reproducing both the SED as well as the VLA and ALMA observations using a physical disk model. We use the D'Alessio Irradiated Accretion Disk (DIAD) code \citep{dal98,dal99,dal01,dal05,dal06}, which has been succesfully used in the past to fit SEDs and (sub)mm images of protoplanetary disks (e.g., \citealp{cal05,hug09,esp11,oso14,mcc16,rub18}). 

DIAD is a $1+1$D code that self-consistently calculates the radial and vertical structure of protoplanetary disks while enforcing hydrostatic equilibrium. The total disk surface density ($\Sigma$) is set mainly by the mass accretion rate in the disk ($\dot{M}$), and the disk viscosity ($\alpha$; $\Sigma\propto\dot{M}/\alpha$), which are free parameters of the model. The dust is distributed in two different populations: small dust grains in the disk atmosphere (at $z>0.1H$, where $H$ is the hydrostatic scale height of the disk), and large dust grains in the midplane (at $z<0.1H$). The grain size distribution of these two populations is assumed to be a power-law, starting at a minimum size of $0.005~\mu$m and with a power of $-3.5$ \citep{mat77}. The maximum grain size (a$^{\rm{atm}}_{\rm{max}}$ in the atmosphere, and a$^{\rm{mid}}_{\rm{max}}$ in the midplane) is also set as a free parameter. 

Dust settling is included by reducing the dust-to-gas mass ratio of the atmospheric dust population, while enhancing it in the disk midplane. The depletion in the atmosphere is parametrized using $\epsilon_{\rm{atm}}$, which is defined as the dust-to-gas mass ratio in the atmosphere relative to the ``standard" dust-to-gas mass ratio. We assume a fixed ``standard" dust-to-gas mass ratio of 0.01. The increase in the dust-to-gas mass ratio in the disk midplane is described by $\epsilon_{\rm{mid}}$, which can be calculated from $\epsilon_{\rm{atm}}$ by enforcing vertical conservation of dust mass at any given radius \citep{dal06}. The radial distribution of $\epsilon_{\rm{mid}}$ can also be modified in order to take into account the effects of dust radial drift, i.e., the increase in the abundance of large dust grains at the inner regions of the disk.
We refer to \citet{dal06} and references therein for a more detailed description of the code.

Our SED model mainly comprises a stellar photosphere and the thermal emission of an irradiated disk (computed with DIAD). In order to account for the direct irradiation of the star onto the edge of the disk cavity, a heated wall is also incorporated \citep{dal05}. Furthermore, we included optically thin dust inside the disk cavity, following \citet{esp11}. As shown by these authors, the stellar photosphere and optically thin dust very well reproduce the optical and near/mid-IR data of GM Aur. Since these two components do not contribute to the mm emission of GM Aur, we chose to add them to our model without any important modification. Thus, we focused on finding the DIAD model parameters that best reproduce GM Aur's SED at wavelengths longer than $\sim10~\mu$m, as well as the VLA and ALMA observations.

\subsubsection{Modeling Procedure}

\begin{figure*}[ht!]
\figurenum{4}
\plottwo{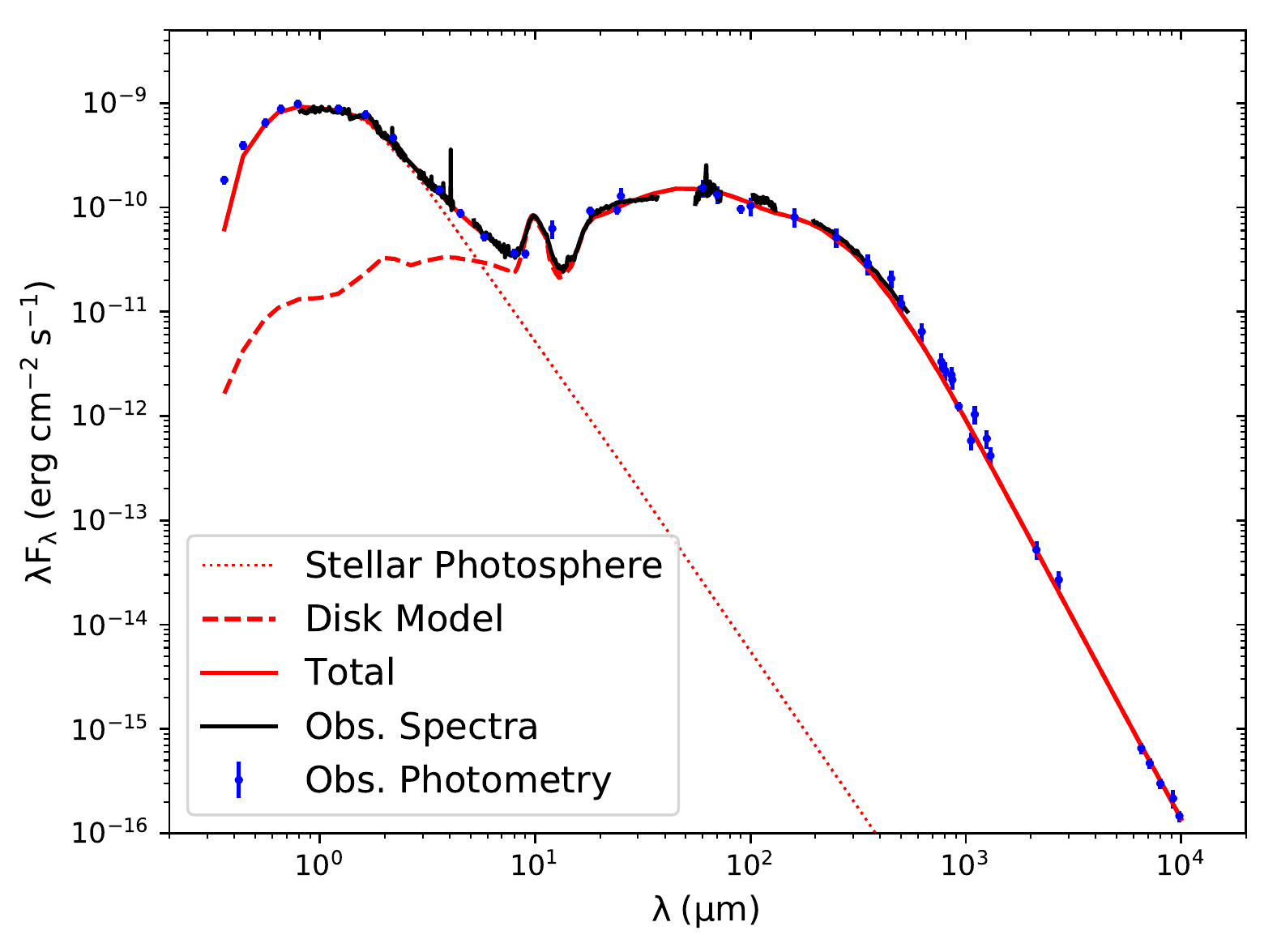}{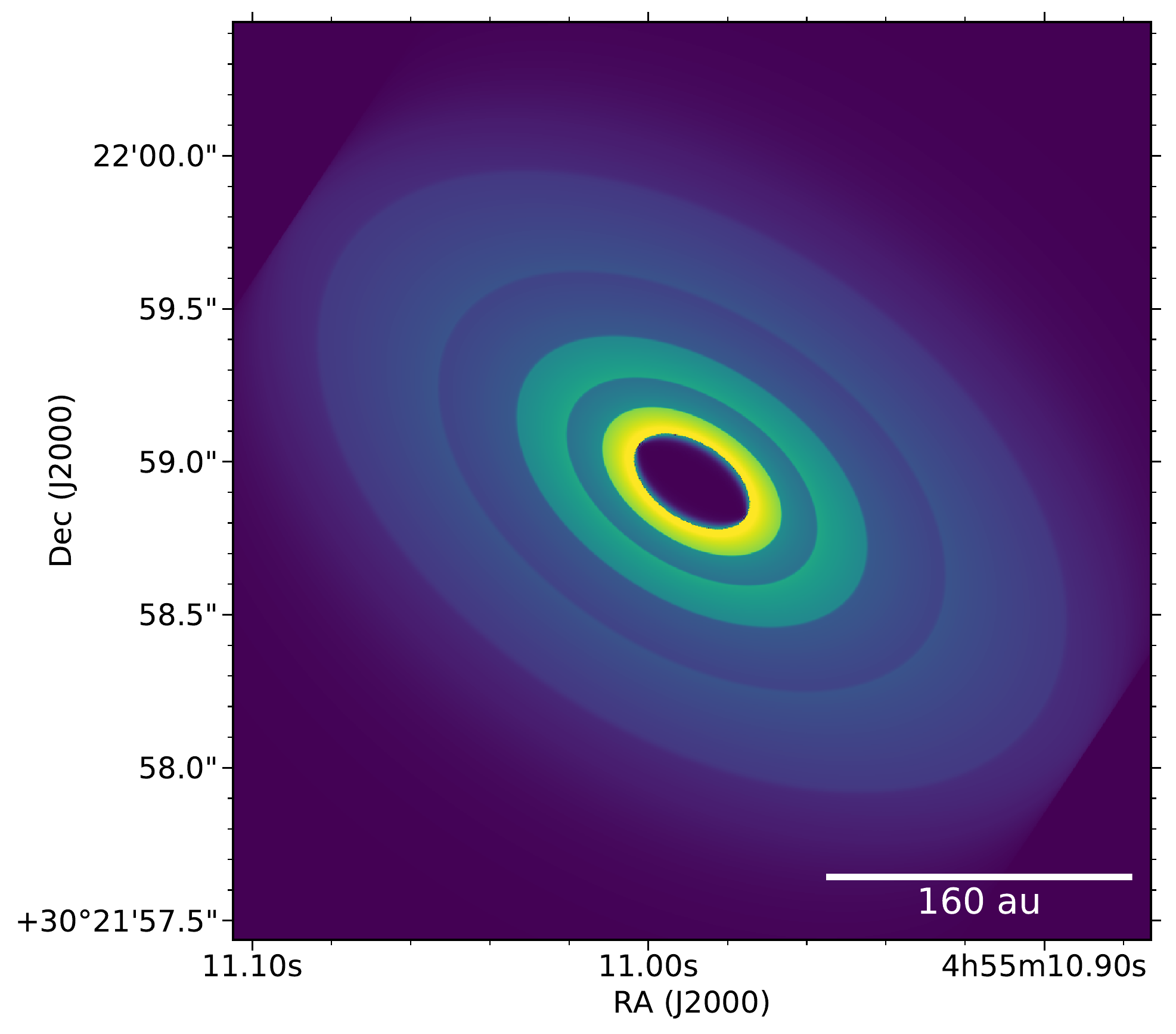}
\label{fig:SED}
\caption{Left: Spectral Energy Distribution (SED) of GM Aur. The blue points and error bars represent the measured photometry, from the optical to cm wavelengths. Black solid lines indicate different observed spectra. The red lines show the different components of our DIAD model (see \S3.2): the dotted line represents the stellar photosphere, the dashed line is the disk model, and the solid line shows the total emission of the model. Right: Synthetic image of the best fit DIAD model at 0.9 mm.}
\end{figure*}

As a first step, we built the SED of GM Aur by compiling photometric and spectroscopic data from the literature, from optical to cm wavelengths (left panel in Fig. \ref{fig:SED}). Most optical to far-IR data were taken from \citet{esp11} and references therein. Additionally, we included Herschel photometry and spectra \citep{rib17}, as well as (sub)mm and cm photometry, from 0.62 mm to 5 cm (\citealp{mac16}, and references therein).

GM Aur's stellar properties were taken from \citet{esp11}. {We note that the recent Gaia Data Release 2 (DR2; \citealp{gai18}) has yielded a distance of $160\pm2$ pc for GM Aur, slightly different to the previously used distance of 140 pc. The stellar luminosity and radius were modified accordingly. We used the stellar isochrones obtained by \citet{sie00} to obtain a stellar mass using the new luminosity and the previously measured effective temperature. We obtained a stellar mass of 1.1 $M_{\odot}$, similar to the value used in previous studies.} The mass accretion rate on GM Aur has been measured to be variable, between $1.1\times10^{-8}$ $M_{\odot}$ yr$^{-1}$ and $3.9\times10^{-9}$ $M_{\odot}$ yr$^{-1}$ \citep{ing15}. $\dot{M}$ and $\alpha$ are significantly degenerate in DIAD, since the disk surface density is mainly set by these two parameters ($\Sigma\propto\dot{M}/\alpha$). Therefore, we decided to fix $\dot{M}$ to an average value of $7.8\times10^{-9}$ $M_{\odot}$ yr$^{-1}$, and we varied $\alpha$ between 0.0001 and 0.01 to find the best-fit model. As a consequence, other models with similar disk masses but different values of $\alpha$ and $\dot{M}$ may also result in satisfactory fits. 

Our VLA and ALMA observations indicate that the distribution of large dust grains in the disk of GM Aur presents a complex radial structure (see \S\ref{subsec:SBPmodel}). We model this substructure by modifying $\epsilon_{\rm{mid}}$ by a factor $\delta$ at different ranges of radii while ensuring that the global dust-to-gas mass ratio of the disk does not depart significantly from our assumed value of 0.01. We note that the disk substructure, and particularly the multiple rings revealed by our surface brightness profile model (see \S\ref{subsec:SBPmodel}), could also be incorporated in a disk model by reducing the total disk density at different radii. However, we choose to only vary $\epsilon_{\rm{mid}}$ since our VLA and ALMA observations only provide information about the large dust grains located in the disk midplane. Gas observations in other protoplanetary disks have shown that gas depletion in disk cavities or gaps is usually lower than for the dust \citep{van15}. Additionally, gas cavities can also be smaller than mm-dust cavities \citep{van16,van18}. Thus, a detailed analysis of molecular line observations would be needed to accurately estimate the substructure of gas in GM Aur, which is outside the scope of this paper.

We use a similar dust composition to \citet{esp11}, who analyzed in detail the \textit{Spitzer} IRS spectrum of GM Aur: $\sim38\%$ of graphite and $\sim62\%$ of silicates, which are in turn composed of $50\%$ of olivine and $50\%$ of pyroxene. We fixed the inclination and position angle of the disk model to the values estimated in \S\ref{subsec:SBPmodel}. The rest of the free parameters of our model were varied in order to reproduce the SED as well as the VLA and ALMA observations. These include the temperature of the disk wall ($T_{\rm{wall}}$), which in turn sets the radius of the disk cavity ($R_{\rm{wall}}$; \citealp{dal05}), the height of the disk wall ($z_{\rm{wall}}$), the maximum grain size in the disk atmosphere (a$^{\rm{atm}}_{\rm{max}}$) and midplane (a$^{\rm{mid}}_{\rm{max}}$), the degree of dust settling ($\epsilon_{\rm{atm}}$), and the disk outer radius ($R_{\rm{disk}}$). Values of $R_{\rm{wall}}$ between 25 and 40 au were explored, with $z_{\rm{wall}}$ being slightly varied around the hydrostatic scale height in order to obtain a good fit to the mid-IR emission. The maximum grain size in the disk atmosphere was varied between the ISM value ($0.25~\mu$m) and $10~\mu$m. For the midplane, larger values between 1 mm and 2.5 cm were explored. The degree of settling was varied around relatively standard values, from 0.001 to 0.8. Finally, we explored values of $R_{\rm{disk}}$ between 150 au and the disk radius estimated from molecular line observations, 300 au.

As a consequence of the complexity of DIAD, computing a disk model is relatively computationally expensive. This prevents us from using common algorithms or statistical methods, such as Levenberg-Marquardt $\chi^2$ minimization or MCMC, to estimate the best-fit parameters of the model and/or their uncertainties. 
Instead, we ran a number of grids of models and found the best-fit parameters by a combination of model selection based on minimum $\chi^2$, visual inspection, and successive refinement of the parameters of the grids. The fact that we are fitting resolved observations of the disk, together with its SED, helps to break several degeneracies that arise when only fitting one of these two observables.

The inspection and selection of models was split in two parts. First, we selected a few models that provided a good fit to the SED of GM Aur while also fitting the 0.9 and 7 mm flux densities. This was done by visually inspecting the models that yielded the minimum $\chi^2$ in the whole grid. Secondly, we obtained synthetic images of these models (right panel in Fig. \ref{fig:SED}) and simulated the VLA and ALMA observations by computing the model visibilities using the same uv coverage as the observations. Additionally, we obtained the residual visibilities by subtracting the model from the observed visibilities, and we produced cleaned images of the model and residuals. We compared our model to the observations using both the deprojected visibilities and the averaged radial intensity profiles. 
After inspecting the results, we made adjustments to $\epsilon_{\rm{mid}}$ and $R_{\rm{wall}}$ in order to obtain a disk structure that produced a good fit to the ALMA and VLA observations. Then, a new grid of models was run while slightly varying the rest of the model parameters. This process was repeated until a good fit was found for the SED while also ensuring that the residuals in the VLA and ALMA images were below $\sim3\sigma$. 

\subsubsection{Model Results}

\begin{figure*}
\figurenum{5}
\plotone{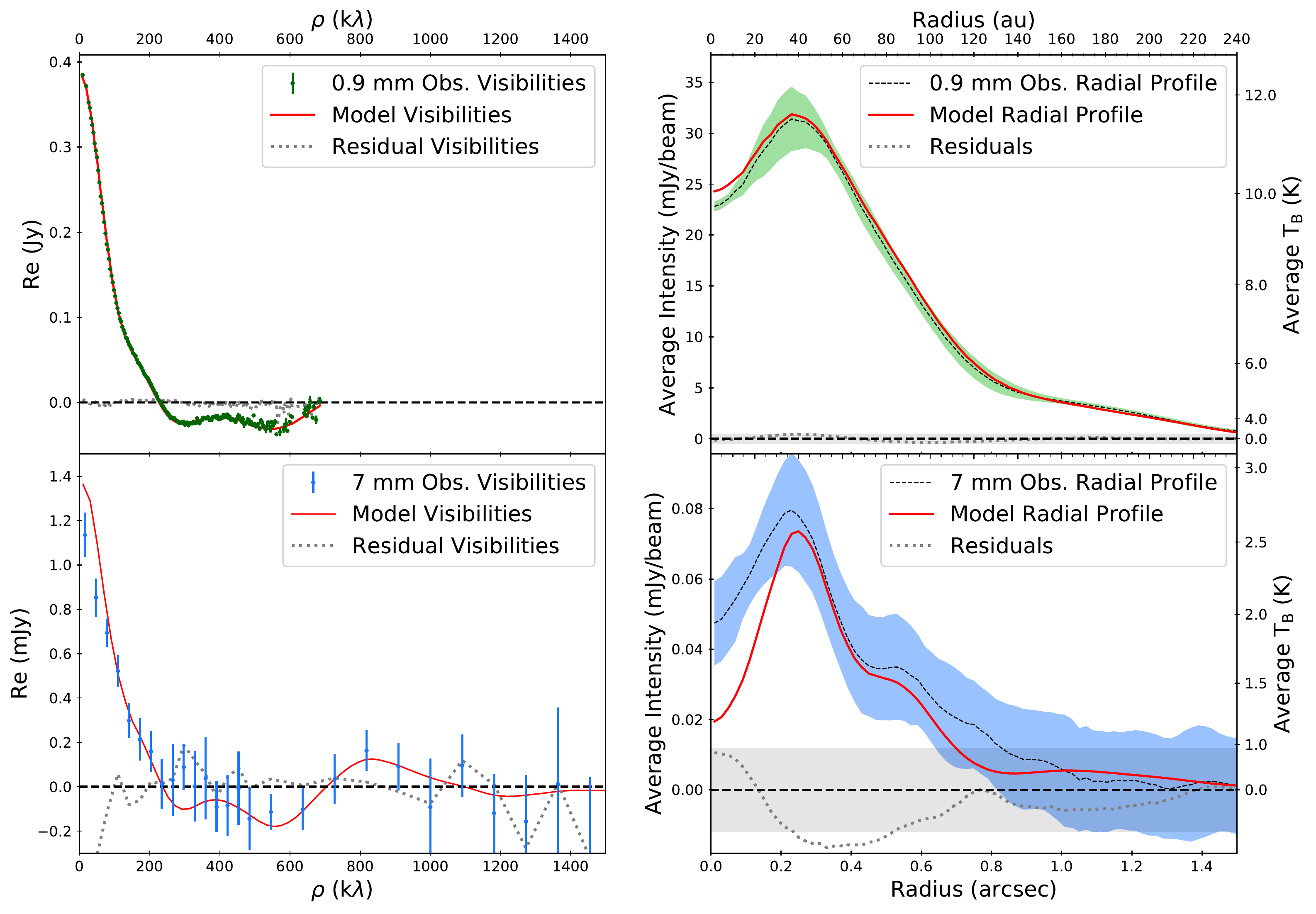}
\label{fig:model_profiles}
\caption{Simulated ALMA (0.9 mm; top panels) and VLA (7 mm; bottom panels) observations computed from our DIAD model of GM Aur. The left panels show the real part of the deprojected interferometric visibilities, while the right panels show the averaged intensity radial profiles obtained from the deconvolved images (Fig. \ref{fig:model_images}). In all panels the observations (green: 0.9 mm, blue: 7 mm), model (red solid lines), and residuals (grey dotted lines) are plotted. The horizontal dashed line indicates the y=0 level. The coloured regions in the right panels indicate the standard deviation at each radii, while the gray shaded region shows the $\pm1\sigma$ (rms of the images) region above and below 0.}
\end{figure*}

\begin{figure*}
\figurenum{6}
\plotone{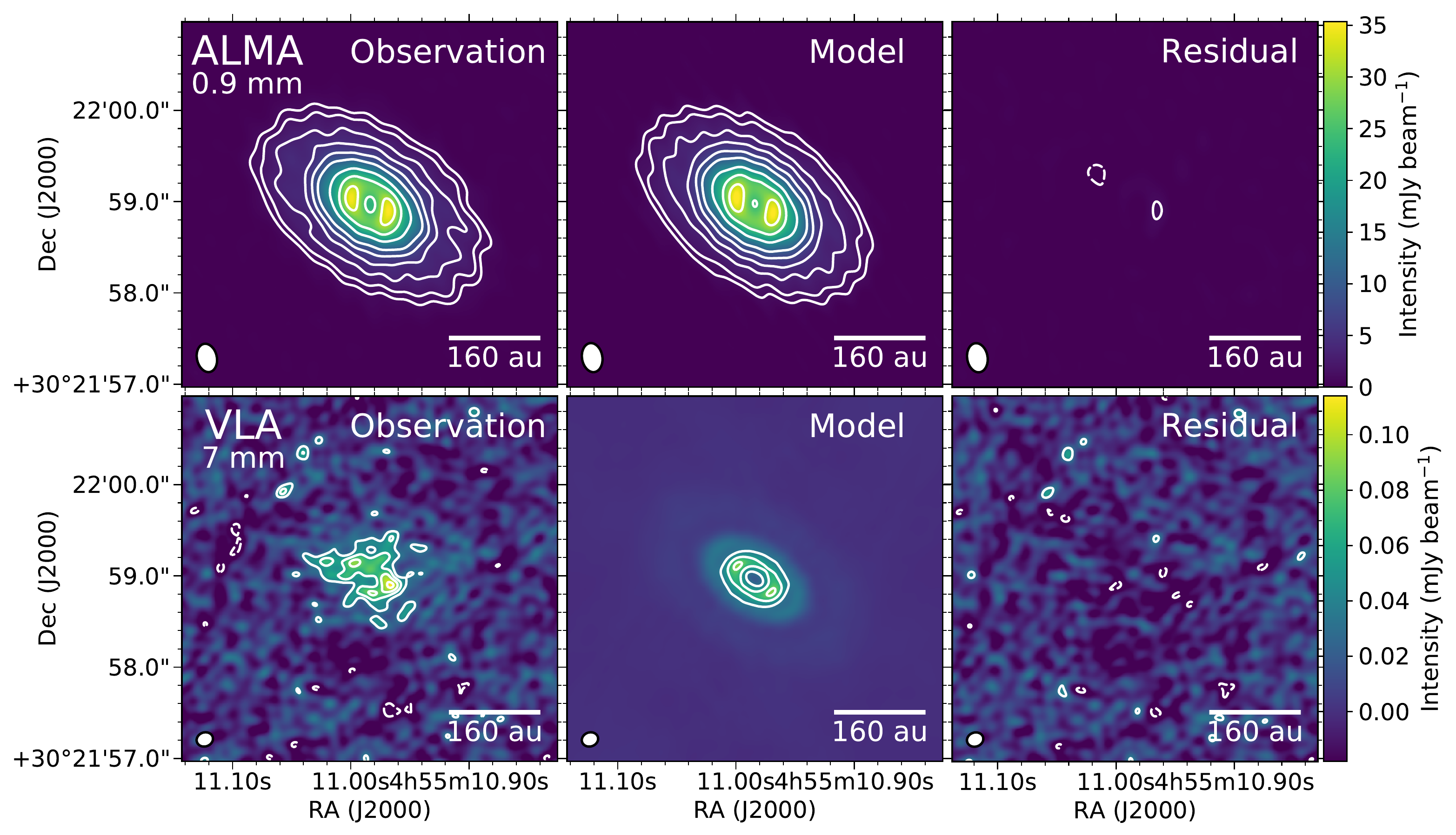}
\label{fig:model_images}
\caption{Comparison between the observations and the simulated images of our DIAD model of GM Aur at 0.9 mm (top panels) and 7 mm (bottom panels). The left panels show the observed images. The middle panels show the simulated observations, obtained by Fourier transforming the images from our model (right panel in Fig. \ref{fig:SED}) with the uv coverage of the observations and then performing the same cleaning process. The residuals (observations - model, computed in the visibility domain) are plotted in the right panels. The contour levels are -3, 3, 5, 9, 13, 20, 30, 50, 70, and 90 times $0.35$ mJy beam$^{-1}$ for the top panels, and $-$3, 3, 5, 7, 9, 11, and 13 times $12~\mu$Jy beam$^{-1}$ for the bottom panels.}
\end{figure*}

The final best-fit parameters of our DIAD model are shown in Table \ref{Tab:model}. In this same table we also show
the factors applied to $\epsilon_{\rm{mid}}$ within the different ranges of radii in order to fit our VLA and ALMA observations. The fits to the SED and real part of the visibilities are in Figures \ref{fig:SED} and \ref{fig:model_profiles}, respectively. A comparison of the observed, model, and residual deconvolved images is shown in Fig. \ref{fig:model_images}. Our best-fit model yields an excellent fit to the SED of GM Aur, as well as to the VLA and ALMA observations.

\floattable
\begin{deluxetable}{lcc}[ht!]
\tablecaption{Stellar Properties and Model Parameters\label{Tab:model}}
\tablehead{
\colhead{Parameter} & \colhead{Value} & \colhead{Reference}
}
\startdata
\multicolumn{3}{c}{Stellar Properties}\\
\hline
$A_V$                             &          0.8         & 1 \\
Spectral Type                     &          K5.5        & 1 \\
$T_{\star}$ (K)                   &          4350        & 1 \\
$L_{\star}$ ($L_{\odot}$)         &          1.2         & 4 \\
$M_{\star}$ ($M_{\odot}$)          &          1.1         & 4 \\
$R_{\star}$ ($R_{\odot}$)          &          1.9         & 4 \\
$\dot{M}$ ($M_{\odot}$ yr$^{-1}$) &   $7.8\times10^{-9}$ & 2 \\
Distance (pc)                     &          160         & 3 \\
\hline
\multicolumn{3}{c}{Disk}\\
\hline
Inclination ($^{\circ}$)          &          52.8        & 4 \\
PA ($^{\circ}$)                   &           56.4         & 4 \\
$T_{\rm{wall}}$ (K)               &         109        & 4 \\
$R_{\rm{wall}}$ (au)              &          34        & 4 \\
$z_{\rm{wall}}$ (au)              &           5.6        & 4 \\
a$^{\rm{atm}}_{\rm{max}}$ ($\mu$m)&            2         & 4 \\
a$^{\rm{mid}}_{\rm{max}}$ (mm)    &           10         & 4 \\
$\epsilon_{\rm{atm}}$             &           0.7        & 4 \\
$\alpha$                          &          0.0012      & 4 \\
$R_{\rm{disk}}$ (au)              &           250        & 4 \\
$M^{\rm{dust}}_{\rm{disk}}$ ($M_{\rm{J}}$)    &           2.0        & 4 \\
\hline
\hline
\multicolumn{3}{c}{$\delta$ Factors for $\epsilon_{\rm{mid}}$} \\
\hline
\multicolumn{2}{c}{Range of Radii} & \colhead{$\delta$} \\
\multicolumn{2}{c}{(au)}  & \colhead{ } \\      
\hline
\multicolumn{2}{c}{0.0  --  34}     & 0    \\
\multicolumn{2}{c}{34  --  52}    & 4    \\
\multicolumn{2}{c}{52  --  73}    & 0.01 \\
\multicolumn{2}{c}{73  --  102}    & 3    \\
\multicolumn{2}{c}{102  --  145}     & 0.01 \\
\multicolumn{2}{c}{145  --  220}      & 1    \\
\multicolumn{2}{c}{220  --  250}      & 0.01\\
\enddata
\tablerefs{(1) \citet{esp10} (2) \citet{ing15} (3) \citet{gai18} (4) This work.}
\end{deluxetable}

By fitting the resolved observations we have been able to estimate the radial structure of large dust grains in the disk, which is a key ingredient to understand the dust evolution and planetary formation process in protoplanetary disks. Our model is formed by two compact rings and a more extended disk. These three components are separated by gaps in the distribution of large dust grains, with a depletion factor of $\gtrsim0.01$. The two inner rings, located at radii $34-52$ au and $73-102$ au, have an enhanced $\epsilon_{\rm{mid}}$ by a factor of 4 and 3, respectively, compared to the expected value from vertical conservation of dust mass. The outer disk component, extending from 145 au to 220 au, has an unmodified $\epsilon_{\rm{mid}}$. At radii $>220$ au, our model predicts that the large dust grains will be significantly depleted, with small dust grains and gas dominating the disk structure. Overall, our model shows a clear increase of the abundance of large dust grains toward the inner regions, combined with dust accumulations in narrow rings. This structure can be explained by the effects of radial drift and dust trapping (see \S\ref{subsec:rings}).

Furthermore, our simultaneous fit to the SED (left panel in Fig. \ref{fig:SED}) allowed us to constrain several physical properties of the disk such as $\epsilon_{\rm{atm}}$, a$^{\rm{atm}}_{\rm{max}}$, a$^{\rm{mid}}_{\rm{max}}$, and $R_{\rm{disk}}$. Interestingly, our model predicts significant dust growth in the disk, with a$^{\rm{atm}}_{\rm{max}}=2~\mu$m and a$^{\rm{mid}}_{\rm{max}}=1$ cm, a sign of a relatively old stage in the dust evolution process in the disk. However, we also find a relatively low degree of settling ($\epsilon=0.7$, so a depletion of only $70\%$ for the small dust grains in the disk atmosphere), which is expected for relatively young disks. Higher degrees of settling (i.e., lower values of $\epsilon$), would predict too much mm emission while also failing to produce enough emission between $\sim200~\mu$m and $\sim600~\mu$m.

The dust mass ($M^{\rm{dust}}_{\rm{disk}}$) in our disk model is $2.0~M_{J}$. Using the dust-to-gas mass ratio of 0.01 assumed in our model, this implies a total disk mass of $200~M_{J}$. Our model is able to constrain very accurately the dust mass of the disk, but since we are only fitting the dust emission, slightly different gas masses could be found by modifying the dust-to-gas mass ratio in the model. We note, however, that since we use a physical model that enforces hydrostatic equilibrium, the temperature structure of the disk, set mainly by the vertical dust distribution, is not entirely independent from the gas density in the disk. 
Therefore, our model is able to obtain a close estimate of the gas density structure of the disk and, hence, of the total disk mass.
Our estimated value is consistent with previous measurements from the literature. In particular, \citet{mcc16} constrained the gas mass of GM Aur to be between 25 and 204 $M_{J}$ by fitting its HD (J=1$-$0) molecular emission with a similar model.

\section{Discussion}\label{sec:discussion}

\subsection{The dust distribution in GM Aur: Radial Drift and Trapping}\label{subsec:rings}

Our 0.9 mm and 7 mm observations have given us the most detailed view so far of the distribution of large dust grains in the disk of GM Aur. The deconvolved images, as well as the surface brightness profile models, indicate that most of the large dust particles in GM Aur are located at a radius $<100$ au. This is confirmed by our physical model, which shows that the dust-to-gas mass ratio of the large dust grains increases toward the central star, even by a factor of 4 at the edge of the inner cavity. However, our model still needs small dust grains extending up to 250 au in order to explain the SED of the disk. As a comparison, the scattered light, which traces the small dust grains in the disk, and the CO emission, which traces the gas, extend up to $\sim300$ au in GM Aur \citep{sch03,hug09}. The morphology of these three disk components (gas, small dust grains, and large dust grains) is a consequence of dust radial drift \citep{whi72}.

Nevertheless, the most striking result of our analysis is that the compact millimeter emission of GM Aur is actually composed of (at least) two rings. Our analytical and physical models clearly show the presence of two bright rings of radii $\sim40$ au and $\sim80$ au, followed by a fainter extended component. While our physical model includes the latter as a third ring from 145 to 220 au, it is uncertain from our analytical model whether this third component should be considered a separate third ring (i.e., whether the disk shows a gap of emission between the second ring and the outermost component). Higher angular resolution observations should be able to discern if this extended component of emission is arising from a separate ring or from a continuous extended disk. 

In any case, our results imply that GM Aur should join the growing group of protoplanetary disks that have been shown to present a multi-ringed structure (e.g., \citealp{oso14,alm15,and16,ise16,mac17,loo17}), increasing the evidence that this type of substructure is common, and maybe even ubiquitous \citep{zha16}. Interestingly, GM Aur presents a multi-ringed structure together with a relatively large extent in its mm emission ($\sim200$ au). This morphology appears to be consistent with the recently proposed idea that large, multi-ringed, primordial disks could be the precursors of transitional disks such as GM Aur, as giant planet formation occurs and clears the inner regions of the disk \citep{gar17,van18}.

Ringed substructures represent a solution for one of the fundamental problems faced by the core accretion scenario: the drift barrier. Theoretical models predict that the drift timescales of mm/cm-sized particles are very short, preventing the dust from growing beyond $\sim$m sizes \citep{bra07}. Local maxima in the gas pressure of the disk can halt or even stop the radial drift of large particles, creating radial dust traps that in turn allow the particles to grow up to larger sizes \citep{pin12}. The results of our physical model are consistent with this scenario: in order to reproduce the ringed morphology of GM Aur, our model needs an increase in the dust-to-gas mass ratio of the large dust grains in the two rings. This dust distribution indicates that the large dust grains in GM Aur are getting trapped and accumulated in two pressure bumps. 

The innermost ring is likely linked to the cavity in the disk. As mentioned above, gaps or cavities cleared by dynamical interactions with planets will produce a pressure peak at the edge of the gap or cavity \citep{zhu14}. Most of the properties of GM Aur indicate that its cavity is probably formed through the interaction with one or more young planets \citep{hug09}, which is also consistent with the results of our physical model. 
It is worth noting that, if one massive planet was responsible for the clearing of the whole cavity in GM Aur, one would expect to have significant departures from axisymmetry such as spiral arms or a vortex (e.g., \citealp{zhu14}). Multiple planets, despite also producing vortices under certain conditions, can result in the clearing of more axisymmetric cavities. If real, the slight asymmetry seen in our observations could be associated with a dynamically induced vortex, but the overall axisymmetry of the observations suggests that the inner cavity of GM Aur has been opened by multiple planets rather than a single massive one.

The origin of the second ring is not as clear. Our results might signal the presence of a planet that is opening a gap between $\sim52$ and $73$ au, but without a robust constraint on the gas depletion inside the gap this scenario cannot be confirmed. Alternatively, a number of mechanisms have been proposed to explain the multi-ringed morphologies recently revealed by ALMA, without the need of invoking several forming planets. These proposed scenarios span a wide range of physical processes, such as the dynamic interaction with massive planets resulting in multiple rings \citep{bae17}, MRI instabilities in magnetized disks \citep{flo15}, changes in dust properties associated to snowlines of volatiles \citep{oku16,pin17b}, or dust traps produced by the back-reaction of the dust onto the gas \citep{gon15,gon17}. 
A better characterization of the dust and gas properties at the second ring will be needed to discern between all these scenarios. 

\subsection{Smaller Disk Cavity at Longer Wavelengths}\label{subsec:cavity}

The ALMA and VLA observations of GM Aur might indicate that the disk cavity has different sizes at 0.9 mm and 7 mm, as shown by our physical model, which used the same disk structure at both wavelengths, and our surface brightness profile model.
The latter finds a smaller radius for the innermost ring at 7 mm ($x_1=39\pm1$ au) than at 0.9 mm ($x_1=42.5\pm0.2$ au).
Additionally, even though the residuals of our physical model (computed in the visibility domain) do not present any significant emission (i.e., $>5\sigma$), the model slightly overpredicts the emission at the center of the disk at 0.9 mm, and falls below the intensity detected at 7 mm (Fig. \ref{fig:model_profiles}). 
These two results suggest that the disk cavity could be different at both wavelengths. 

If this discrepancy is confirmed, it could be the result of a different spatial distribution of the dust grains traced at each wavelength. The continuum emission at 7 mm probes larger dust particles than at 0.9 mm. 
Thus, the fact that the cavity looks smaller at 7 mm could indicate that the mm/cm-sized dust grains are more accumulated at the cavity edge than the smaller grains traced at 0.9 mm. As mentioned above, large dust grains are expected to get trapped and accumulate at the pressure bump located at the edge of a cavity like the one present in GM Aur. Such a pressure bump is expected to be radially asymmetric\citep{zhu12,zhu14}. Thus, since the gas drag will be stronger on mm/cm-sized particles than on submm particles, dust grains traced at 7 mm are expected to be more tightly accumulated at the edge of the cavity (e.g., \citealp{zhu12,flo15,pin15}).
Due to the limited angular resolution of the observations, such a dust distribution could result in an apparent shift of the peak of the ring at 7 mm toward smaller radii. At the same time, this effect should result in narrower rings at longer wavelengths. Unfortunately, our surface brightness profile model does not constrain the width of the ring as well as its position, so we cannot accurately evaluate whether there is a significant difference in the ring width at 0.9 mm ($\sigma_1=0\rlap.''047\pm0\rlap.''002$) and 7 mm ($\sigma_1=0\rlap.''044\pm0\rlap.''014$).

Another possibility is that the difference in cavity size is related to the presence of ionized gas inside the disk cavity. Free-free emission from ionized gas will likely be located close to the central star. This could produce an increase in the 7 mm emission at the inner regions of the disk, which would result in an apparent smaller cavity. Based on cm observations of GM Aur, $\sim15\%$ of the 7 mm emission is expected to be free-free emission, either from an ionized jet or from photoionized gas \citep{mac16}. This observational evidence makes free-free emission the most plausible scenario to explain the difference in cavity size. 

With our current data we cannot rule out any of the two proposed scenarios, but new observations could easily shed some light. Higher sensitivity observations at 7 mm could confirm whether the cavity has different sizes at 0.9 mm and 7 mm. These data, together with high angular observations at cm wavelengths, could help us characterize the free-free emission and find out whether it can explain the difference in cavity sizes. Furthermore, high angular resolution ALMA observations at intermediate wavelengths ($\sim1-3$ mm) will allow us to test whether the larger dust grains in the disk tend to accumulate closer to the cavity edge by resolving spatial changes in the spectral index of the (sub)mm emission.

\section{Summary and Conclusions}\label{sec:conclusion}

We have presented ALMA (0.9 mm) and VLA (7 mm) observations of the dust continuum emission of the protoplanetary disk around GM Aur. The deconvolved images at both wavelengths clearly resolve the inner cavity of the disk. Additionally, the 0.9 mm image displays an extended and fainter disk component that is not evident at 7 mm. 

We have modeled the observed visibilities at these two wavelengths following two different approaches: an analytical model of the surface brightness profile, and a physical model of the disk that also reproduces the SED of GM Aur.

The results of our model have revealed that the disk of GM Aur is composed of at least two bright rings of dust, supporting the idea that rings and other types of substructures are common in protoplanetary disks. Our physical model shows that the observed ringed morphology can be explained by an enhancement in the abundance of large dust grains in two concentric rings. These enhancements are probably caused by gas pressure bumps that are trapping the mm/cm-sized dust particles of the disk. The inner pressure bump is likely produced by the dynamical clearing of the inner cavity, while the second ring could be produced by different physical mechanisms.

Finally, the inner cavity appears to be smaller at 7 mm than at 0.9 mm. This discrepancy could be indicating that the mm/cm-sized dust grains, traced at longer wavelengths, are more accumulated at the cavity edge than the submm-sized particles. Another possibility is that the free-free emission at 7 mm is producing an apparent shift of the cavity edge toward inner radii. New observations are needed to confirm this discrepancy and shed some light on its possible origin.

\acknowledgments

\vspace{5mm}
EM, CCE, and AR acknowledge support from the National Science Foundation under CAREER Grant Number AST-1455042 and the Sloan Foundation. G.A., M.O., and J.F.G. are supported by the MINECO (Spain) AYA2014-57369-C3 and AYA2017-84390-C2 grants (co-funded by FEDER).

This paper makes use of the following ALMA data: ADS/JAO.ALMA\#2016.1.00565.S. ALMA is a partnership of ESO (representing its member states), NSF (USA) and NINS (Japan), together with NRC (Canada), MOST and ASIAA (Taiwan), and KASI (Republic of Korea), in cooperation with the Republic of Chile. The Joint ALMA Observatory is operated by ESO, AUI/NRAO and NAOJ.
This work has made use of data from the European Space Agency (ESA) mission {\it Gaia} (\url{https://www.cosmos.esa.int/gaia}), processed by the {\it Gaia} Data Processing and Analysis Consortium (DPAC, \url{https://www.cosmos.esa.int/web/gaia/dpac/consortium}). Funding for the DPAC has been provided by national institutions, in particular the institutions participating in the {\it Gaia} Multilateral Agreement. 

This paper utilizes the \citet{dal98,dal99,dal01,dal05,dal06} Irradiated Accretion Disk (DIAD) code.  We wish to recognize the work of Paola D'Alessio, who passed away in 2013. Her legacy and pioneering work live on through her substantial contributions to the field.



\facilities{VLA, ALMA}

\appendix
\section{Markov Chain Monte Carlo runs}\label{appendix:MCMC}
We have analyzed our VLA and ALMA observations of GM Aur by modeling the profile of the real part of the deprojected visibilities using an analytical model of the surface brightness profile. Information about the model and its results can be found in \S\ref{subsec:SBPmodel}. Here we describe some details about the Markov Chain Monte Carlo (MCMC) run used to obtain the posterior distributions of the model parameters.

\begin{figure*}
\figurenum{A1}
\centering
\includegraphics[height=0.9\textheight]{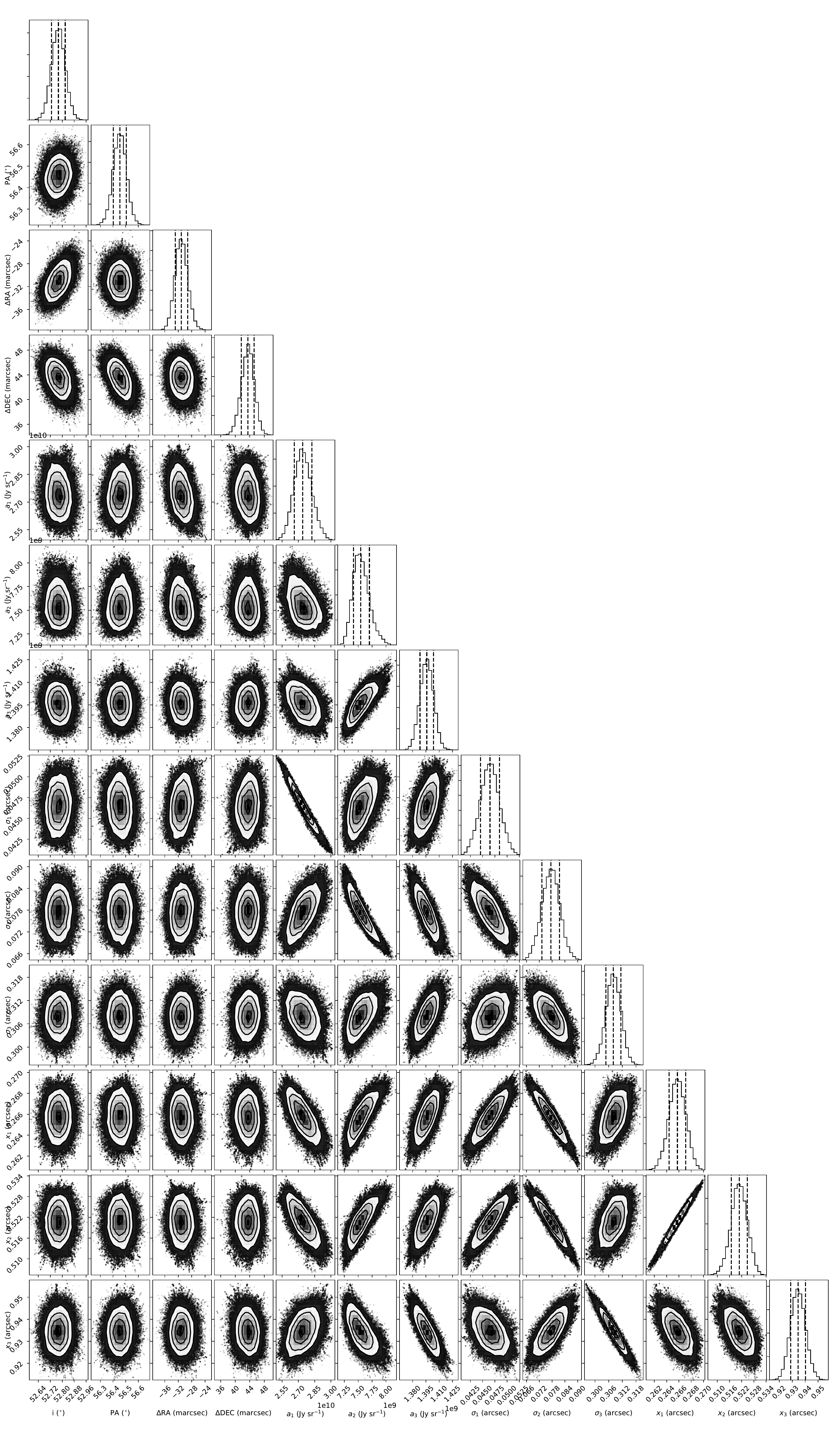}
\label{fig:cornerALMA}
\caption{Corner plot showing the posterior distributions for the parameters of our surface brightness profile model of the ALMA observations at 0.9 mm. The vertical dashed lines indicate, from left to right, the 16th percentile, median, and 84th percentile of each posterior.}
\end{figure*}

We set the initial positions of the chains around the best-fit solution obtained using a Levenberg-Marquardt $\chi^2$ minimization algorithm. The priors for the fit to the ALMA data were set flat between reasonable values. These priors allowed a proper exploration of the parameter space, without setting the limits for any of the posteriors (see Fig.\ref{fig:cornerALMA}). 

Since the 0.9 mm emission is detected with such a high signal-to-noise ratio, we used the results from the fit to the ALMA observations to motivate the priors of the inclination, PA, standard deviation ($\sigma_i$), and position ($x_i$) of the Gaussian rings used for the VLA visibilities. We did this by using Gaussian priors centered at the median of the posteriors obtained by ALMA. The standard deviations of the Gaussian priors for the inclination and PA were set to the standard deviation of the ALMA posteriors. For $\sigma_i$ and $x_i$ we used more flexible priors with 10 and 50 times the standard deviation of the ALMA posteriors, respectively. Furthermore, we used a Gaussian prior for the RA and DEC offsets, with its standard deviation set to the VLA absolute position accuracy ($\sim10\%$ of the beam size). Flat priors were used for the scale factors of the Gaussian rings ($a_i$). These priors helped to obtain a better fit to the VLA visibilities without fully setting the posterior distributions of $\sigma_i$ and $x_i$, as can be seen from the significantly different results found at the two wavelengths (see Table \ref{Tab:SBPmodel}).

\begin{figure*}
\figurenum{A2}
\plotone{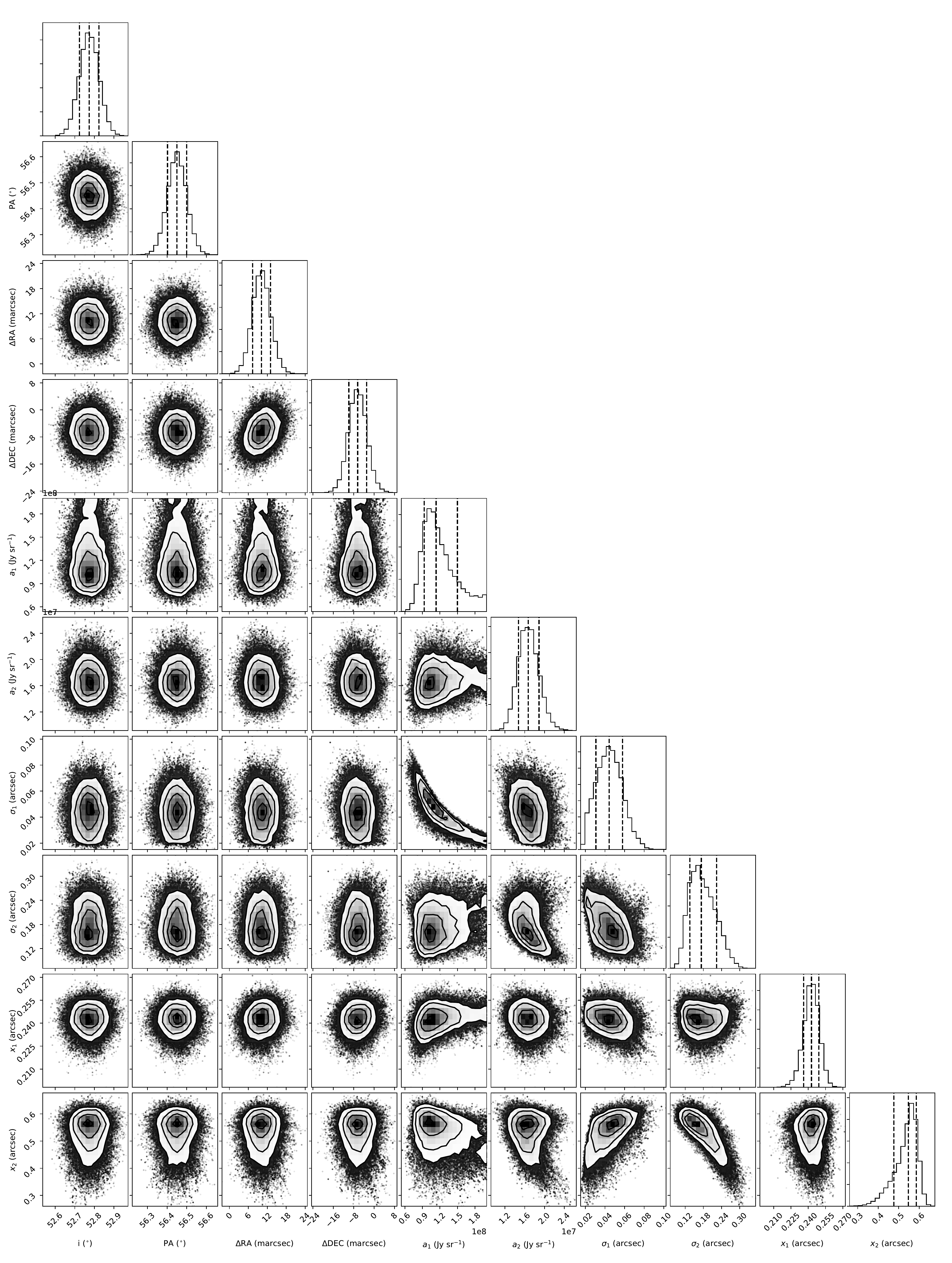}
\label{fig:cornerVLA}
\caption{Corner plot showing the posterior distributions for the parameters of our surface brightness profile model of the VLA observations at 7 mm. The vertical dashed lines indicate, from left to right, the 16th percentile, median, and 84th percentile of each posterior.}
\end{figure*}

We ran the MCMC for 30000 iterations to assure that the chains had converged. The last 5000 iterations were used to obtain the posteriors of all the model parameters. The obtained corner plots for the ALMA and VLA observations are shown in Fig.\ref{fig:cornerALMA} and Fig.\ref{fig:cornerVLA}, respectively.

\clearpage


\listofchanges

\end{document}